\pdfoutput=1

\documentclass[11pt]{article}
\usepackage{ifthen} 
\newboolean{showcomments}
\setboolean{showcomments}{true} 
\ifthenelse{\boolean{showcomments}}
   {

}

\usepackage[]{ACL2023}
\usepackage{tabulary}
\usepackage{multirow}
\usepackage{graphicx}
\usepackage{float}
\usepackage{times}
\usepackage{latexsym}
\usepackage{url}
\usepackage{array}
\newcolumntype{P}[1]{>{\centering\arraybackslash}p{#1}}
\usepackage[T1]{fontenc}

\usepackage[utf8]{inputenc}

\usepackage{microtype}

\usepackage{inconsolata}

\title{Understanding the Challenges of Deploying Live-Traceability Solutions}

 \author{Katherine R. Dearstyne \and Alberto Daniel Rodriguez \and Jane Cleland-Huang \\
        University of Notre Dame \\ Notre Dame, IN\\ \texttt{\{kdearsty, arodri39, janeclelandhuang\}@nd.edu}}
\begin{document}
\maketitle
\begin{abstract}
Software traceability is the process of establishing and maintaining relationships between artifacts in a software system. This process is crucial to many engineering processes, particularly for safety critical projects; however, it is labor-intensive and error-prone. Automated traceability has been a long awaited tool for project managers of these systems, and due to the semantic similarities between linked artifacts, NLP techniques, such as transformer models, may be leveraged to accomplish this task. SAFA.ai is a startup focusing on fine-tuning project-specific models that deliver automated traceability in a near real-time environment. The following paper describes the challenges that characterize commercializing software traceability and highlights possible future directions.
\end{abstract}
\begin{table*}[!t]
    \centering
    \begin{tabular}{|P{2.5cm}|p{9.5cm}|P{1.75cm}|P{.75cm}|}
        \hline
        \textbf{Name (domain)} & \textbf{Description} & \textbf{Candidates Links} & \textbf{True Links} \\ \hline
        
        CM1 (Embedded System) & Requirements for a data processing unit for the then NASA metrics data program. Contains high (22) and low level requirements(53) and prepared by researchers at the University of Kentucky \cite{hayes_cm1}. & 1,166 & 45 \\ \hline
        
        Medical Infusion Pump (Healthcare) & Software system implementing a medical infusion pump extracted from \cite{Larson2013_mip}. Contains trace links between system components (21) and regulatory requirements (126). & 2,778 & 132 \\ \hline
        
        iTrust (Healthcare) & An electronic health record (EHR) system created as part of a course at North Carolina State University \cite{Meneely2011_itrust}. Contains requirements (131) and JSP code modules (226). & 29,606 & 534 \\ \hline
        
        Dronology (UAV) & A system for managing the navigation of UAVs and their communication to the ground control station. Contains requirements (55), designs (99), and java code (458). Prepared by researchers at the University of Notre Dame \cite{clelang_huang_dronology}. &  &  \\ 

        D-NL & Subset containing only requirements to designs. & 5,445 & 58 \\ 

        D-PL & Subset containing only designs and java code. & 45,342 & 232 \\ \hline
        
        
    \end{tabular}
    \caption{Descriptions of the datasets evaluated in study. Candidate links represent all the potential combinations between the source and target artifacts. True links represent the number of those combinations that are positively linked, the remaining ones are considered unlinked. All datasets were extracted from coest.org}
    \label{tab:datasets}
\end{table*}

\section{Introduction}
\label{sec:intro}
Software traceability is a critical task for many software systems and entails linking high-level software artifacts (e.g. requirements or safety-goals) to their fulfillment, often in the form of design definitions, source code or test-cases \cite{Gotel_Finkelstein_1994, FOSE, torkar_requirements_2012}. By creating accurate trace links in a project, engineers can analyze how the introduction of a change in the system might impact existing components \cite{Hamilton_Beeby_1991}. These links also aid in verifying the completeness of a project and provide clarity onto the rationale of an artifact's inception \cite{Ramesh_Edwards_1992}. Specifically, trace links might demonstrate that all requirements have been fulfilled or highlight which requirements are addressed by specific design decisions. In many safety-critical systems, traceability is required to assure their safety by governing bodies.

Unfortunately, creating and maintaining trace links is an effort-intensive, time-consuming, error-prone task which may impede the development process \cite{hayes_advancing_2006}. As a result, many projects are left with incomplete and inaccurate trace links \cite{mahmoud_semantic_2012}. Since most linked artifacts share semantic similarities, natural language processing techniques may be leveraged to support engineers in identifying missing trace links, reducing both the time and cost required for the task \cite{Antoniol_Recovering_2002, dekhtyar_technique_2007, guo_tackling_2017, Lin_Liu_Zeng_Jiang_Cleland_Huang_2021}.

\begin{figure}
    \centering
    \includegraphics[width=.9\columnwidth]{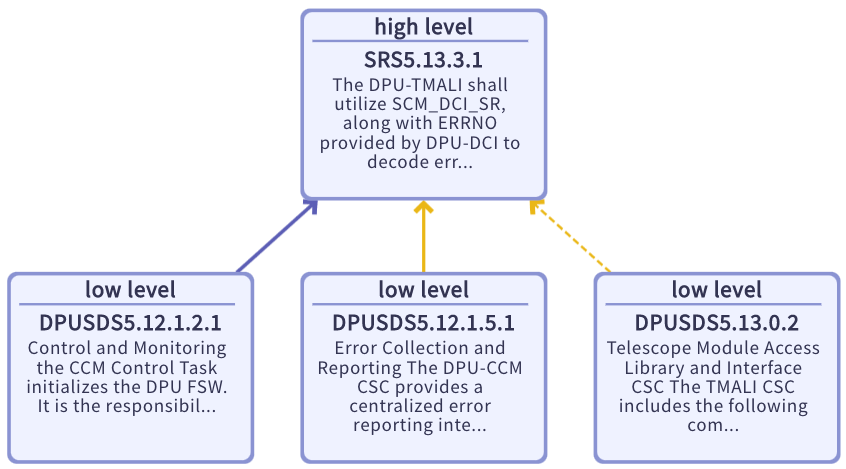}
    \caption{Example sub-tree from the SAFA platform, linking high-level requirements to low-level requirements in the CM1 dataset. The solid blue line is an \textbf{established} relationship as determined by the project engineers. The yellow lines are relationships \textbf{predicted} by SAFA, where the dotted line represents a prediction that has not yet been manually approved while the solid line is an approved link.}
    \label{fig:cm1_safa}
\end{figure}

Inspired by this goal, SAFA (``Software Artifact Forest Analysis'') \cite{rodriguez_safa_2022} is an emerging traceability tool that was based on recent research showing the achievement of BERT and other bi-directional transformers at predicting trace links. 
\cite{Lin_Liu_Zeng_Jiang_Cleland_Huang_2021}. Despite the promising results of these models in the research setting, we have encountered challenges in deploying SAFA on commercial projects. These challenges have included the lack of data-availability, necessity of domain-specific knowledge, poor quality of many datasets, and expectations of immediate results from engineers. Furthermore, many safety-critical domains require near-perfect recall when using NLP techniques to automatically generate links, which can not be consistently achieved by the current models. Therefore, for the focus of this paper, we explore each of these problems in depth within the area of traceability and highlight insights into future directions. 

Although there remain challenges that must still be overcome, SAFA shows promise in bringing the vision of ``ubiquitous traceability'' to fulfillment at last \cite{FOSE}. In this vision, traceability will occur automatically alongside the engineering processes, incurring little cost and allowing engineers to focus on those tasks that are most significant to them. By integrating the latest advances in NLP into the platform, SAFA can reduce the burden of traceability and accelerate engineering processes.

\begin{table*}[!t]
    \centering
    \begin{tabular}{|P{3cm}|P{3cm}|p{9cm}|}
        \hline
        \textbf{Name} & \textbf{Model Id} & \textbf{Description}\\ \hline
        nl-bert & \emph{ANONYMOUS/nl-bert} & Bert-base-uncased model pre-trained on tracing commit messages to related issues. Traces between two sets of natural language artifacts. \cite{Lin_Liu_Zeng_Jiang_Cleland_Huang_2021}. \\ \hline
        pl-bert  & \emph{ANONYMOUS/pl-bert} & Robert-base model pre-trained on docstring to code snippets and git commit to commit code. Traces between natural language and programming language artifacts \\ \hline
        bert-base-uncased  & \emph{bert-base-uncased} & The original BERT model released by Google in 2019 \cite{Devlin_Chang_Lee_Toutanova_2019}.  \\ \hline
        VSM &  N/A & Uses the scikit-learn Tf-Idf vectorizer and related utilities for calculating similarity scores \cite{scikit_learn}. \\ \hline
    \end{tabular}
    \caption{Description of deep learning and classical models used in evaluation. Model id is the name of the model in the HuggingFace model repository (\emph{https://huggingface.co}).}
    \label{tab:models}
\end{table*}

\section{NLP for Automated Traceability}
\label{sec:related}
There has long been a need for developer tools focusing on the traceability of software artifacts \cite{Hamilton_Beeby_1991}, and this need has only grown as these systems have become more complex, distributed, and collaborative. Previous approaches to automated traceability included vector and topic based information retrieval techniques such as vector-space model (VSM), latent semantic indexing (LSI), and latent dirichlet allocation (LDA) \cite{Antoniol_Recovering_2002, Asuncion_Asuncion_Taylor_2010}. However, these approaches have many short-comings. For example, VSM fails to correctly link artifacts that relate to one through synonyms rather than identical terms, referred to as the \emph{term-mismatch problem} \cite{Guo_Gibiec_Cleland_Huang_2017}. Although topic based techniques can overcome instances of the term-mismatch problem by translating artifacts into their latent space, this translation looses some of the necessary information required for tracing, resulting in VSM generally out-performing these approaches \cite{Hamilton_Beeby_1991, Asuncion_Asuncion_Taylor_2010}. Interestingly, VSM and topic-based techniques were shown to bring orthogonal information about the trace links as combining their similarity scores helped to mitigate their individual problems \cite{Gethers_Oliveto_Poshyvanyk_Lucia_2011}. Nevertheless, performance still fell short of the accuracy needed for commercial applications, and it became evident that a model would have to understand synonyms, the  relationships between the words in a sentence, and be tuned to best understand the specific vocabulary of a project. Classical statistical methods were not equipped to overcome these challenges; however, deep learning models, like word2vec \cite{Mikolov_Chen_Corrado_Dean_2013}, re-invigorated the community by showing that models could create contextualized embedding for sentences, hinting that it could be possible to capture all this information.

The advances in deep learning caused researchers to re-think what was possible for software traceability. First, the use of RNNs (recurrent neural networks) and Bi-GRU (Bidirectional Gated Recurrent Unit) were explored and shown to outperform previous baselines like VSM or LSI \cite{Guo_Cheng_Cleland_Huang_2017}. As exciting as these results were, these models were data hungry and required large amounts of data to be able to out-perform the baselines. Finally, the introduction of transformer based models, such as Google's BERT model and OpenAI's GPT models, resulted in an additional performance leap, showing the potential for breaking through the previous glass ceiling \cite{Feng_Guo_Tang_Duan_Feng_Gong_Shou_Qin_Liu_Jiang, Lin_Liu_Zeng_Jiang_Cleland_Huang_2021, Lin_Poudel_Yu_Zeng_Jiang_Cleland_Huang_2022}. One particular strength of these models was their ability to be fine-tuned to specific projects and domains, finally enabling achievement of highly accurate results. \cite{Lin_Liu_Zeng_Jiang_Cleland_Huang_2021, Lin_Poudel_Yu_Zeng_Jiang_Cleland_Huang_2022}. This is the focus of SAFA, fine-tuning state-of-the-art traceability models to create a real-time project-management environment with live-traceability \cite{rodriguez_safa_2022}.

\section{Current Performance Demonstration}
\label{sec:experiment}

\begin{table}[!t]
    \centering
    \begin{tabular}{|P{1.1cm}|P{2.1cm}|P{.7cm}|P{.7cm}|P{.9cm}|}
        \hline
        \textbf{Dataset} & \textbf{Model} & \textbf{MAP} & \textbf{F2} & \textbf{Train Time} \\ \hline
        CM1 & VSM & 71.4 & 46.4 & \textbf{<1s} \\ 
        CM1 & BERT-BASE & 60.8 & 45.8 & 12.1m \\
        CM1 & NL-BERT & \textbf{72.3} & \textbf{57.2} & 12.2m \\ \hline

        MIP & VSM & \textbf{100} & 38.9 & \textbf{<1s}\\ 
        MIP & BERT-BASE & \textbf{100} & \textbf{100} & 28m \\ 
        MIP & NL-BERT & \textbf{100} & \textbf{100} & 27.7m \\ \hline
        
        D-NL & VSM & 78.0 & 58.7 & \textbf{<1s} \\ 
        D-NL & BERT-BASE & 68.4 & 51.8 & 55.5m\\ 
        D-NL & NL-BERT & \textbf{86.5} & \textbf{64.9} & 55.9m  \\ \hline

        D-PL & VSM & 21.6 & 14.4 & \textbf{<1s} \\ 
        D-PL & BERT-BASE & 39.4 & 39.2 & 8h \\
        D-PL & PL-BERT & \textbf{51.6} & \textbf{46.7} & 8.5h \\ \hline
        
        iTrust & VSM & 28.4 & 24.7 & \textbf{<1s} \\ 
        iTrust & BERT-BASE & 72.1 & 65.6 & 4.5h \\ 
        iTrust & PL-BERT & \textbf{78.7} & \textbf{69.3} & 4.7h \\ \hline
    \end{tabular}
    \caption{The average performance across three random seeds for each of models on all dataset. These results are calculated on the evaluation dataset.}
    \label{tab:results}
\end{table}
In order to demonstrate the current performance of NLP techniques across traceability datasets, we utilize 5 different datasets, spanning four domains as shown in Table \ref{tab:datasets}. To these datasets, we apply VSM \cite{salton_vector_1975}, an un-pretrain bert-base-uncased model \cite{devlin_bert_2019} and one of two pre-trained bert models \cite{Lin_Liu_Zeng_Jiang_Cleland_Huang_2021, Lin_Poudel_Yu_Zeng_Jiang_Cleland_Huang_2022} depending on whether the dataset trace between natural language artifacts or between natural language artifacts and source code. A detailed description of each model can be found in Table \ref{tab:models}.

All candidate travel links in dataset are randomly split into three parts. 35\% of the data was used for training (train), 10\% was used for validation (e.g. performing early stopping), and 55\% was used for the final evaluation. We run each dataset-model combination across three different random seeds to obtain three unique combinations of the data splits. We trained all transformer models for 20 epochs and with a batch size of 4, performing gradient accumulation for 16 steps. All the models produce similarity scores between source-target artifact combinations ranging from 0 to 1. We selected to train with less than 50\% of the data to simulate project's that need to produce trace links during the development process.

To evaluate model performance, we calculate the Mean Average Precision (MAP) and max F2 scores as described in \cite{Lin_Poudel_Yu_Zeng_Jiang_Cleland_Huang_2022}. These metrics are commonly used in the traceability community and were selected because MAP helps favor the precision of the model while F2 favors recall. All results are displayed in Table \ref{tab:results}, and an example of the pre-trained model predictions in the SAFA platform is shown in Figure \ref{fig:cm1_safa}.

\begin{figure*}
    \centering
    \includegraphics[width=.85\textwidth]{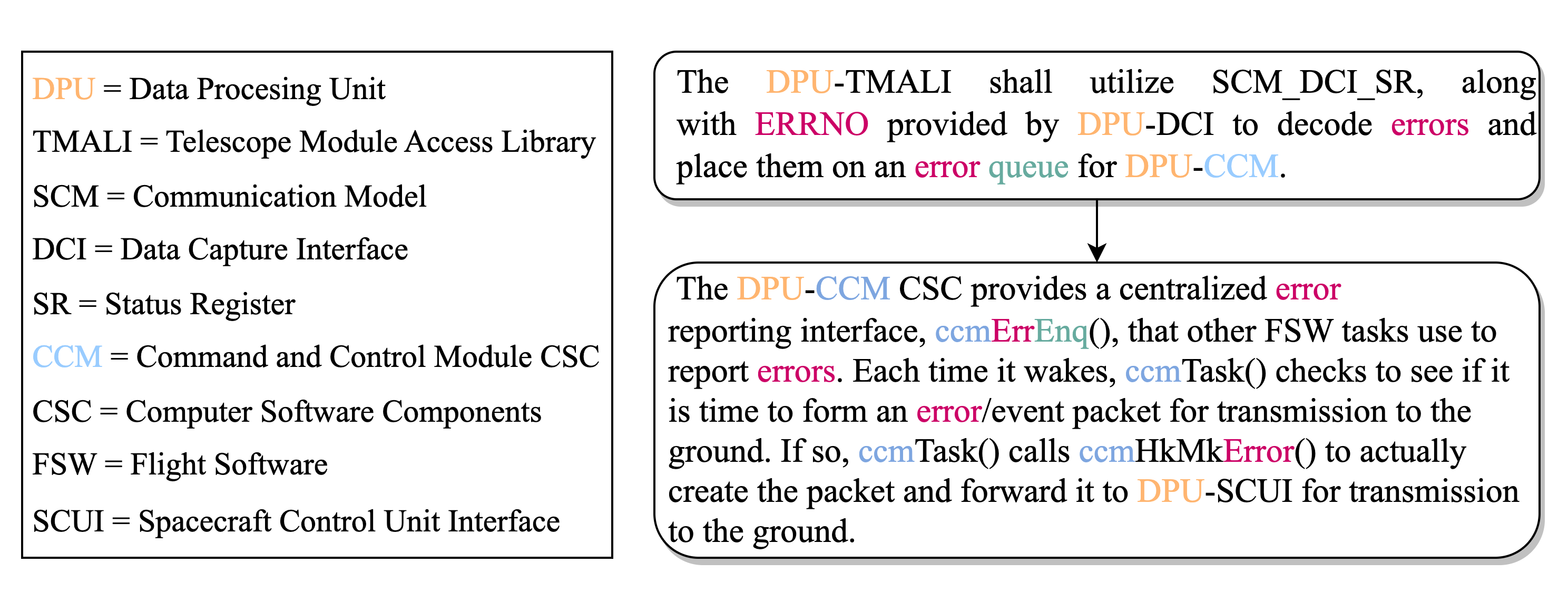}
    \caption{Explanation of trace link from CM1 slice.}
    \label{fig:explanation}
\end{figure*}  

\section{Problems Encountered}
\label{sec:problems}
\subsection{Data-Availability}  
\label{sec:prob_data}
As with many research areas within software engineering, automated traceability suffers from the lack of available data to train models. The ideal dataset would contain a complete software project across multiple layers of artifacts with high quality trace links, but few such datasets are open source. This problem is only intensified when focusing on safety-critical systems, where accurate traceability is often required by regulatory bodies. Due to protecting proprietary data, many potential industry partners are only willing to provide example projects under the condition that the data remains on their premises, requiring that all models be trained using the companies own computing resources. This unfortunately limits the resources which can be dedicated to model training and predicting, inhibiting a traceability tool from being able to guarantee real-time results as described in Section \ref{sec:prob_time}.  Furthermore, many industry projects in non safety-critical domains do not contain any trace links at all due to the additional time required to trace artifacts throughout the development process. However, these companies often desire traceability after project completion to validate that all requirements have been met. This presented a challenge as fine-tuning models on example links from the target project is important in allowing the models to reach satisfactory performance levels \cite{Lin_Liu_Zeng_Jiang_Cleland_Huang_2021}. Previous research has achieved some degree of success by performing transfer learning using commits linked to issues in git-repositories, allowing the model to better learn the task of traceability \cite{Lin_Poudel_Yu_Zeng_Jiang_Cleland_Huang_2022}. However, the performance of such models in a few- or zero-shot task is highly variable and can not reach the level of performance of the fine-tuned models.

One possible approach to increase the available links for fine-tuning is to use VSM to make predictions over a subset of the data. Since VSM is able to obtain competitive MAP scores on many natural language datasets, its predictions would likely capture a significant number of true trace links. Afterwards, an engineer could manually review predictions with the highest similarity scores, and the resulting trace-links could be fed to the model as training data, hopefully boosting model performance on the remaining links. We plan to explore this approach with a pilot customer in the upcoming months.

Recently, the NLP community has seen great improvements in few-shot learning problems, as demonstrated by GPT3 \cite{brown_language_2020}. There have also been some encouraging examples of few-shot learning within requirements engineering \cite{alhoshan_zero-shot_2023}. In the future, we are eager to see how these recent advancements might improve trace link predictions in projects where there is insufficient training data. 

\subsection{Domain-Specific Knowledge}
\label{sec:prob_domain}
Since requirements and other software artifacts tend to include highly technical, domain-specific jargon, models pre-trained on only a general corpus of text are often unable to perform well on traceability tasks \cite{Lin_Liu_Zeng_Jiang_Cleland_Huang_2021}. For this reason, domain-specific pre-training and/or transfer-learning are especially important to these models.

Due to the similarity of git-hub repositories to the task of tracing, domain-specific projects make particularly good transfer-learning material. This motivated us to explore open-source repositories within one of the domains most in need of traceability, namely robotics. Unfortunately, our search turned up only 8,790 public repositories on the topic of 'robotics'. This contrasted starkly with Keras, one of the datasets on which the model was shown to perform best  \cite{Lin_Poudel_Yu_Zeng_Jiang_Cleland_Huang_2022} Keras included tags for `data-science' and `machine-learning', which each contain over 30K search results. This suggests that supplemental training strategies and data may be required for domains that are not as well-represented by open-source public repositories.

To better highlight the problem of domain-specific vocabulary, we provide an examination of a trace link (Figure \ref{fig:explanation}) from the data slice of CM1 presented in the Section \ref{sec:intro}. The high level requirement details that the \emph{DPU-TMALI}, a telescope module access library, should utilize the status register, \emph{SCM\_DCI\_SR}, to decode errors and place them on a queue for another component, \emph{DPU-CCM}. The lower level requirement describes how the \emph{DPU-CCM} module checks its queue for errors and how those errors are forwarded to the control unit, \emph{DPU-SCUI}, before being sent to the ground station. The explanation was generated by reading the specification for the requirements \cite{noauthor_software_2003} and took about one hour for a knowledgeable traceability researcher to sufficiently understand. While constructing or vetting trace links would likely take domain-experts far less time, and be simpler in less-complex projects, this example highlights the complexity of reviewing a single candidate trace link. In future work, we plan on exploring how to make this process more efficient by leveraging different trace link explanation techniques based on either knowledge graphs \cite{Liu_Lin_Zeng_Jiang_Cleland_Huang_2020}, GradCAM equivalents \cite{grad_cam_bert}, or through interactive visualizations of the model's attention \cite{vig_deconstructing_2022}.

\subsection{Training and Prediction Time}  \label{sec:prob_time}
As mentioned earlier, many companies require that models are trained on the companies' own resources so as to ensure that the data is kept private. This limits the amount of computing resources which might be dedicated to the task, thus emphasizing the importance of efficient training and prediction times for a model. Previously, Google's bert-base-uncased showed the most promise for the traceability task \cite{Lin_Poudel_Yu_Zeng_Jiang_Cleland_Huang_2022}, but a model of this size can take up to days to train on large industrial-sized datasets. For example, it took an average of 8 hours to train a bert-base model with 4 NVIDIA T4 GPUs on the largest dataset (cf.~Section \ref{sec:experiment} for more details). Although much of the pre-training and transfer-learning can be completed prior to handing off a model to a company, the fine-tuning and prediction times can still be quite lengthy and many companies expect immediate results. 

Interestingly, VSM, which scales linearly with an increasing number of trace-links, is able to achieve high MAP scores for many natural language datasets. Due to its efficiency, there may be benefits to using VSM in domains where high accuracy is less critical and fast-predictions are prioritized. 

In addition, using knowledge distillation or smaller transformer architectures has been shown to reduce required computational resources while still providing quality results \cite{tay_efficient_2022}. We are actively exploring how these can be applied to traceability.

\begin{figure}[!t]
    \centering
    \includegraphics[width=\columnwidth]{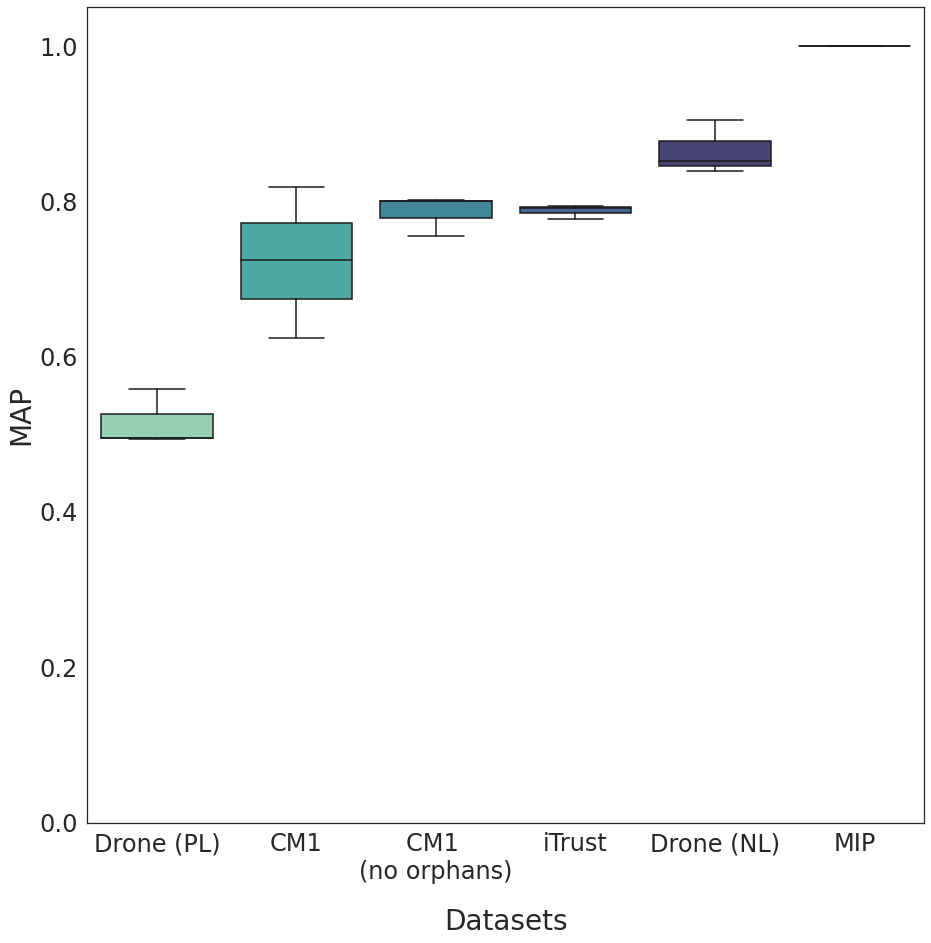}
    \caption{The range of scores across three random seeds for each dataset.}
    \label{fig:variability}
\end{figure}  

\subsection{Model Performance}
\label{sec:prob_performance}
Although models have been shown to reach high F2 and MAP scores in some instances, the results are highly variable across datasets. For example, within our experiment results in Table \ref{tab:results}, there is a difference of 48.4\% MAP between the lowest (D-PL) and highest (MIP) performing datasets . This has several likely root causes, including the lack of quality trace links and the domain specific jargon, as discussed in Sections \ref{sec:prob_data}-\ref{sec:prob_domain}. Additionally, there is a variable level of quality across datasets, and automated traceability can be a difficult, if not impossible, task on lower quality datasets. For example, requirements that are ambiguous or source code that contains poorly named variables and methods can cause a model to fail. Furthermore, during fine-tuning of the model, all artifacts which are not explicitly linked in the project are assumed to be negative links (not traced). In reality, however, these may be true, yet missing links, and a large number of mis-labeled negative labels may prohibit the model from being able to learn correct patterns in the data.

We can clearly see the effects of mis-classified negative labels on the results of CM1. Across random splits of the data, CM1 has the most significant variance with regard to the pre-trained model performance, ranging from an MAP as low as 63\% all the way up to 82\%. It also under-performs in comparison to all other datasets aside from D-PL. Upon further inspection, we found 26 artifacts in the dataset which had no ground truth links, hereby referred to as orphan artifacts. Since the expectation of the CM1 project is for all higher-level requirements to be fulfilled by lower-level requirements, orphan artifacts are likely a sign of forgotten links. When we removed them, performance was boosted to 79\% MAP. Noteably, this new average is quite close to the highest MAP previously seen across random seeds. We speculate that this variance was due to the number of orphans falling into the training set, where a large number of mis-labeled links made it difficult for the model to pick up on meaningful patterns. Indeed, the run of the model without orphans, resulted in more consistency across dataset splits as can be seen in Figure \ref{fig:variability}. Inspired by this finding, we also discovered that D-PL, the dataset with the lowest MAP, had over 300 orphan artifacts. Due to time constraints, we were unable to re-run this experiment on the dataset, but we plan to explore how this might improve performance in the future. 

Although we can detect some forgotten links in the dataset by identifying orphan artifacts, there are likely other missing links that might be challenging to identify. We are presently investigating how we might automatically prune links that are likely mis-labeled \cite{pleiss_identifying_2020} or make the models more invariant to noise \cite{abdar_review_2021}.


\section{Conclusions}
\label{sec:conclusions}
Clearly, automated traceability continues to present challenges that must be faced during the commercialization of the SAFA platform. Nevertheless, there are many potential advancements that might be made by employing state-of-the-art NLP techniques, some of which have been identified in this paper. By collaborating with researchers within NLP, automated traceability tools, such as SAFA, have the potential to reach new levels of performance. Ultimately, this may facilitate a transformation of engineering processes into the long-awaited ideal of ubiquitous traceability.

\section*{Limitations}
\label{sec:limitations}

The main limitation of our paper is the low number of datasets used throughout our evaluations. This was primarily chosen due to time constraints, but limited data availability is a problem as described Section \ref{sec:prob_data}. To improve the breadth of our analysis, we attempted to use projects with diverse artifacts across four domains. The artifact types encompass requirements, design definitions, use cases, java code, and jsb code. However, to make more definite conclusions, we hope to utilize additional data in future works. 

Further limitations include the the lack of reproducibility of our experiment, as our code is intellectual property of SAFA and cannot be released to the public. The datasets used are, however, open source and can be found at \url{http://coest.org}.

\section*{Ethics Statement}
\label{sec:ethics}
Ethical considerations for the lack of open source code must consider that successful commercialization projects are a viable path for impacting the current practice of safety critical systems. As SAFA is still a pre-seed company, our go-to-market plan does not (at least initially) include open sourcing our code base. We plan on releasing different domain-specific models that the community can benefit from. Currently, we have published two models used in this paper \url{https://huggingface.co/ANONYMOUS/nl-bert} and \url{https://huggingface.co/ANONYMOUS/pl-bert}.


\bibliography{anthology,custom}
\bibliographystyle{acl_natbib}

\end{document}